\documentstyle [12pt]
{article}
\begin{document}

\hyphenation {never-theless}
\hyphenation {direct-ly}
\hyphenation {ac-ting}
\hyphenation {desing-ularisation}

\title{Constraints on discrete symmetries from anomaly
cancellation in compactified superstring theories}
\author{Graham G. Ross\thanks{SERC Advanced Fellow}, Christoph
M.A. Scheich\thanks{Supported by the
European Community, Science Program},
\\Department of Physics,\\
Theoretical Physics,\\
University of Oxford,\\
1 Keble Road,\\
Oxford OX1 5DX}
\maketitle
\begin{abstract}
\noindent
Compactified string theories give rise to discrete symmetries
which are essential if they are to provide a realistic low energy
theory. We find that in a class of four dimensional string
theories these symmetries are constrained by similar conditions
to those discrete anomaly cancellation conditions found in the
case the discrete symmetry is a residue of a spontaneously broken
gauge symmetry.  Such conditions strongly constrain the allowed
form of the low energy effective theory.
\end{abstract}
\maketitle

While four dimensional string theories appear to have all the
necessary ingredients needed to give the structure of the
Standard Model, progress towards the construction of realistic
string theory models has been hampered by the embarrassingly
large number of candidate theories. Given this, any generic
property of a string theory which may be used to limit the
effective low-energy theory without committing the analysis to
a specific example is important.

In this letter we explore the possibility that the discrete
symmetries that follow from compactified string theories are
constrained by conditions similar to that were found
\cite{ibanez} to result from anomaly cancellation when a local
gauge symmetry is broken leaving a discrete symmetry factor
unbroken \cite{priv}. The importance of these conditions stems
from the fact
that they involve only the states left light after the breaking
of the gauge symmetry and thus constrain the low energy theory
on its own. The power of these constraints is well illustrated
by the study of discrete gauge symmetries in supersymmetric
versions of the Standard Model where it was shown \cite{ross+}
that only two non-R symmetry discrete groups of low dimension
satisfy the discrete anomaly cancellation conditions while
suppressing nucleon decay giving just two candidate versions for
the supersymmetric Standard Model, the minimal version which
conserves B and L and a variant violating L but not B.

An advantage of a discrete gauge symmetry is that the constraints
following from such a symmetry are not violated by gravitational
corrections \cite{krauss}, an extremely important consideration
if such constraints are needed to stabilise the nucleon for
example. While in what follows we will be only able to show that
some of the discrete non-R symmetries following from
compactified string theories satisfy discrete anomaly
cancellation conditions it is suggestive that, if they do, they
too are discrete gauge symmetries and protected from large
gravitational corrections.
The string theories we consider here are all based on the
heterotic string compactification. The main part of the paper is
concerned with the 4D-construction of Gepner, although we will
briefly comment on the implications for associated Calabi-Yau
constructions.

 The Gepner models are $D$-dimensional ($D<10$, even) heterotic
string models in which compactification is achieved by tensoring
minimal $N=2$ superconformal models to form an internal sector
of the theory in such a way that  conformal anomaly
cancellation is achieved and a $N=1$ space-time supersymmetric,
modular invariant theory with correct spin-statistics is
generated \cite{gepner}. These models are characterized by their
level $k_i$ and their type according to the $ADE$-classification
scheme  \cite{ade}.
They have conformal anomaly
\begin{equation}
c_i\ =\ \frac{3k_i}{k_i+2}\ \ .
\end{equation}
The states of the theory are formed from tensor products of the
primary fields
$\Phi^{l;\overline{l}}_{q,s;\overline{q},\overline{s}}$
associated with each factor. The conformal dimensions and
$U(1)$-charges of these fields are given by
\begin{equation}
h\ =\ \frac{l(l+2)-q^2}{4(k+2)}\ +\ \frac{s^2}{8}
\end{equation}
\begin{equation}
Q\ =\ -\frac{q}{k+2}\ +\ \frac{s}{2}\ \ .
\end{equation}
The heterotic string theory then possesses in general a maximal gauge
symmetry $E_{D/2+4}\otimes E_8'\otimes U(1)^{r-1}$.
The full set of discrete symmetries with generators in the
left-moving sector may be written as a $Z_{k_i+2}\otimes Z_2$ symmetry
with charges $\bar{q}_i^J,\; \bar{s}_i^J$ satisfying
\cite{andy,cordes}
\begin{eqnarray}
\sum_{J=1}^{n}\bar{q}_{i}^{J} =0 & mod\ k_i+2 &
(\forall i)
\label{s1}\\
\sum_{J=1}^n \bar{s}_i^J =0 & mod\ 2 &
(\forall i)\ .
\label{s11}
\end{eqnarray}
The right moving part of the string has a
$\tilde{Z}_{k_i+2}\otimes \tilde{Z}_2$ symmetry which is
generically an R symmetry with charges $q_i,\; s_i$ satisfying
\cite{andy,cordes}
\begin{eqnarray}
\sum_{J=1}^n q_i^J + 2 = 0 & mod\ k_i+2 &
(\forall i)
\label{s2}
\\
\sum_{J=1}^n s_i^J + 2\Sigma_{J=4}^n d_i^J = 0 & mod\ 2 &
(\forall i)
\label{s22}
\end{eqnarray}
and follows from the selection rules for non-vanishing
correlation functions involving $n$ fields which come from the
parafermionic nature of the primary fields \cite{kopp}.
Here $d_i^J$ describes the vertex operators in the $(0)$ picture.
We have adopted the convention that the quantum numbers $q$, $s$
refer to the fields in the $(-1)$ picture and the $\bar{q}$,
$\bar{s}$ refer to the
appropriate choice of representation leading to a gauge invariant
correlation function.

It will be important, in what follows, to identify which discrete
symmetries are discrete subgroups of the original gauge
symmetries. In the left moving part of the string the $U(1)$
gauge group of each tensor factor $i$ contains a discrete
subgroup  $Z_{2(k_i+2)}$ if $k_i+2$ is odd or $Z_{k_i+2}$ if
$k_i+2$ is even. The charges, $\bar{q'}^{J}$, of the fields
entering in a non-vanishing correlation function are constrained
by these symmetries to satisfy
\begin{eqnarray}
\sum_{J=1}^n \bar{q'}_i^{J} \equiv
\left\{ \begin{array}{lll}
\sum_{J=1}^n
[-2\bar{q}_i^J + (k_i+2)\bar{s}_i^J] = 0 & mod\ 2(k_i+2) &  \rm
for\ \it k_i\rm +2 \rm\ odd\it
\\
\sum_{J=1}^n [-\bar{q}_i^J + \frac{k_i+2}{2}\bar{s}_i^J] = 0
                                      & mod\ (k_i+2) &
\rm for\ \it k_i\rm +2 \rm\ even\it\ \ (\forall i )\ ,
\end{array} \right.
\label{neulad}
\end{eqnarray}
where $n$ is the number of fields appearing in the correlation
function \cite{andy,cordes} and $i$ refers to the i$^th$ tensor
factor.
These symmetries are a subgroup of the symmetries of
eq.(\ref{s1}),(\ref{s11}).

For all modular invariants the conformal theory is related to a
Calabi-Yau theory \cite{vafa}. The diagonal sum $(q_i+\bar{q}_i)/2$
are the Gepner charges defined for the $\underline{10}_{scalar}$
representation. The associated symmetry is that found in the
Calabi-Yau construction \cite{wir}. Due to correlations between
left and right sectors the discrete group may be somewhat larger
than that of eqs.(\ref{s1})-(\ref{s22}). In fact the diagonal
sum and difference of eqs.(\ref{s1})  and
(\ref{s2}) is a $Z_{2(k_i+2)}\otimes {Z}_{2(k_i+2)}$ symmetry for
$A$, $D_{odd}$, and $E_6$ invariants and a $Z_{k_i+2}\otimes
{Z}_{k_i+2}$ symmetry for $D_{even}$, $E_7$ and $E_8$ invariants
while for (\ref{s11}) and (\ref{s22}) it is a $Z_4\otimes Z_4$
symmetry and a $Z_2\otimes Z_2$ symmetry for the two classes of
invariants respectively \cite{cordes}. However, as we will
discuss, the larger symmetries are all trivially realised on the
spectrum or R symmetries; the non-R symmetries of interest here
are all contained in eqs.(\ref{s1}) and (\ref{s2}).

In the following let us start with a study of the $\bar{q'}$,
$q$, $\bar{q}$,
$s$, $\bar{s}$ charges separately and then apply the result to
linear combinations of the left and right moving charges.
We consider the case that the discrete symmetries associated with
these charges are not spontaneously broken although the
underlying $U(1)$s may be. The $\bar{q}^{\prime}$ charges
automatically fulfil the discrete anomaly cancellation relations
$i$ to $iv$ of \cite{ibanez} since the $Z_{2(k_i+2)}$ or
$Z_{k_i+2}$ symmetry is a subgroup of the $U(1)$s.
Except for an embedding into $E_{D/2+4}$ and $E'_8$ simultaneously
leading to an $U(1)_A$ contained partially in one and the
other, all gauge symmetries are anomaly free \cite{lt}. In the
case of the anomalous $U(1)_A$ the discrete Green-Schwarz
mechanism is at work \cite{ibanezneu} implying the required
cancellation through the axion field.

However (for $k_1+2$ even) the $\bar{q}$, $\bar{s}$ charges are
associated with a larger discrete symmetry and need not satisfy
the discrete anomaly cancellation conditions. Even so, we can
deduce constraints on these charges for the light states from the
connection of these symmetries with the $U(1)^{r}$ gauge
symmetries of the Gepner construction. The latter are all anomaly
free and so the ($M$) massless states of the theory satisfy
anomaly cancellation conditions:

\bigskip

i) Mixed $Z_N$-gravitational conditions.

For the $U(1)$ factors we have the usual mixed $U(1)$
gravitational anomaly cancellation conditions

\begin{equation}
\sum_{J=1}^{M}\alpha_{\bar{q}^{J}_{i},\bar{s}^{J}_{i}}=0
\hspace{1cm} \forall i \ \in\ 1,...,r\ ,
\label{ac}
\end{equation}
where
\begin{equation}
\alpha_{\bar{q},\bar{s}}=[k(k+2)]^{-\frac{1}{2}}[-
\bar{q}+\frac{1}{2}(k+2)\bar{s}] .
\label{alpha}
\end{equation}
Using eq.(\ref{alpha}) in eq.(\ref{ac})
\begin{equation}
\sum_{J=1}^{M}[-
\bar{q}_{i}^{J}+\frac{1}{2}(k_{i}+2)\bar{s}_{i}^{J}]=0\ .
\label{an1}
\end{equation}
we immediately derive the associated  mixed $Z_N$-gravitational
conditions
\begin{equation}
\sum_{J=1}^M \bar{q}_i^J = pN + \eta_q\frac{N}{2}\ ,\ p,q\in Z,
\label{nur}
\end{equation}
where $\eta_q$=1,0 for $N=k_i+2$ even, odd, following from the
fact that $\sum_{J=1}^M\bar{q}_i$ has to be an integer.
These are the usual discrete gravitational anomaly cancellation
conditions, $i$ of \cite{ibanez}.

Using the same arguments it is now straightforward to derive the
remaining anomaly cancellation conditions for the discrete
symmetries in the same basis as above:

\bigskip

ii) Pure discrete $Z_NZ_MZ_L$ cancellation conditions
\begin{equation}
\sum_{i}\ q_ip_io_i\ =\ tN+sM+rL+\eta_qu\frac{N}{2}
                                +\eta_pv\frac{M}{2}
                                +\eta_ow\frac{L}{2}
                      \ \ ,\ \ t,s,r,u,v,w\ \in\ Z\ ,
\label{eq:u13}
\end{equation}
where $\eta_q$, $\eta_p$, $\eta_o$ = $1,0$ for $N$, $M$, $L$
even, odd and $q_i$, $p_i$, $o_i$ are the discrete charges
($=\bar{q}_i$).
Note that this is weaker than the result $ii$ of \cite{ibanez}
in which the last three terms vanish if any one of the discrete
groups is odd. We will discuss below the conditions under which
the two results coincide.

\bigskip

iii) Mixed discrete and gauge cancellation conditions
\begin{equation}
\sum_i\ T_iq_i\ =\
\frac{1}{2} r''N\ +\ s''\eta_q\frac{N}{4}
\ \ , \ \ r'',s''\ \in\ Z\ \ ,
\end{equation}
where $T(R)$ is the quadratic Casimir corresponding to each given
representation $R$ (the normalization is such that the Casimir
for the $M$-plet of $SU(M)$ is $\frac{1}{2}$).
Note that this differs from the result $iii$ of \cite{ibanez} due
to the appearance of the last term.

\bigskip

iv) Mixed discrete and $U(1)$ conditions;  $Z_NU(1)_XU(1)_Y$ and
$Z_NZ_MU(1)_X$
\hfill \break
\begin{eqnarray}
\sum_i\ x_iy_iq_i\ =\ r'''N\ +\ s'''\eta_q\frac{N}{2}\ \ , \ \
r''',s'''\ \in\ Z\ \ ;    \label{least}\\
\sum_i\ x_iq_ip_i\ =\ s^{iv}N+r^{iv}M+\eta_qt^{iv}\frac{N}{2}
                   +\eta_pu^{iv}\frac{M}{2}
                \ \ ,\ \ r^{iv},s^{iv},t^{iv},u^{iv}\
\in\ Z\ \ , \label{last}
\end{eqnarray}
where $x_i$, $y_i$ denote the $U(1)$ charges. Again the last two
terms are in general different from the result $iv$ in
\cite{ibanez}.

Of course, starting with only a knowledge of the discrete charges
of the light states in an effective low energy theory it is
always possible to identify these charges with a {\it larger}
discrete symmetry in such a way as to satisfy these mixed
conditions ({\it i.e.} the charges $x^{\prime}_i=2x_i \; ;
q^{\prime}_i=2q_i$  will satisfy eqs.(\ref{least}) and
(\ref{last}) for
$N^{\prime}=2N$.). As with the discrete gauge anomalies these
mixed conditions are only useful if we restrict the size of the
discrete gauge group and assume the heavy states have
conventional $U(1)$ charges.

In certain cases the differences between the conditions derived
here and the discrete anomaly cancellation conditions disappear.
This happens if the original $E_6$ gauge symmetry is not broken
by nontrivial embeddings of twists. The fields that can
contribute to the discrete anomaly are the matter fields
transforming as singlets under $E_6$ or as the 27 dimensional
representations.
For the case of $E_6$ singlets or the fields that transform as
the ($\underline{1}+\underline{10}$)
representations under the $SO(10)$ subgroup of $E_6$
(components of the 27 dimensional representation of $E_6$) the
$\bar{s}$ are even. For the case of the 16 dimensional
representations of $SO(10)$ the $\bar{s}$ are odd but the
multiplicity is even. Thus in both cases the contribution to the
second term of eq.(\ref{an1}) is an integer multiple of
$(k_{i}+2)$. Hence we derive the discrete
anomaly cancellation condition $i$ without the $\eta$ term. The
same applies for $ii$ to $iv$. However in the case that the
original $E_6$ gauge symmetry is broken by twists nontrivially
embedded (see below for details), it is easy to show that there
can be contributions with $\bar{s}$ odd and odd multiplicity. In
this case the new conditions derived above apply.

So far we have discussed only the anomaly cancellation conditions
associated with the $Z_{k+2}$ symmetries of the $\bar{q}$
charges. The $\bar{s}$ charges also fulfil
eqs.(\ref{nur})-(\ref{last}), but in this case the symmetry is
$Z_2$ and the relations are trivial. The anomaly cancellation
condition for the $\tilde{Z}_{N}$ symmetries associated with the
right sector charges (cf. eqs.(\ref{s2}) and (\ref{s22})) follow
from these results by using the Gepner construction of massless
states as we now discuss.

Consider first the case of untwisted models. One starts with a
combination of half integer charge $Q_{tot}$ and appropriate
conformal dimension $h$ in the right internal sector and obtains
the ones in the left sector by adding ($n$) multiples of
$\beta_0$, and ($m$) multiples of $\beta_i$ \cite{gepner} in such
a way that the resulting states have again half integer
$\overline{Q}_{tot}$ and appropriate $\overline{h}$. Here
$\beta_0$ is the generator of supersymmetry and acts by adding
1 to each $q_i$ and $s_i$ component. The vector $\beta_i$ only
acts by adding 2 to the $i$th $s_i$ index. Thus
\begin{equation}
q_i^J=\bar{q}_i^J -n\ \ \  ,\ \ \
s_i^J=\bar{s}_i^J -n-2m_i\ \  (\forall i)\ .
\label{lr}
\end{equation}
Let us first consider those charges, $q_i^{\prime}$, related by
eq.(\ref{lr}) to the charges $\bar{q}_i^{\prime}$. It is
convenient to define a commuting basis which projects the non-R
symmetries by choosing the linear combination of the $Z_{k_i+2}$
symmetries generated by $\Sigma_{i=1,r}v_iq_i^{\prime}$, {\it
i.e.} labelled by the vector $(v_1,\cdots ,v_r)$, such that
\begin{eqnarray}
\sum_{i=1}^r \frac{k_i}{k_i+2} v_i^pv_i^q = Z\ \ \forall\ p,q\ .
\end{eqnarray}
This basis contains the R symmetry $\beta_0$. The remaining ones
which commute with it are non-R symmetries. It is clear from
eq.(\ref{lr}) that under these non-R symmetries the massless
states have $q^{\prime}$ charges which differ only by multiples
of $2(k+2)$ or $k+2$, for the two cases of eq.(\ref{neulad}),
from the $\bar{q}^{\prime}$. Therefore all these symmetries
fulfil the discrete anomaly gauge cancellation conditions $i$ to
$iv$ of ref. \cite{ibanez} {\it without} the additional terms of
eqs.(\ref{nur})-(\ref{last}).

Of course, the $q^{\prime}$ charges in general do not generate
the complete set of symmetries in the right sector. The full set
of  symmetries with $q_i$ charges may be treated in a similar
manner although in this case the appropriate basis needed to
project the non-R symmetries is given by orthogonal combinations
of the vectors $(v_1,\cdots ,v_r)$ (again containing the R
symmetry $\beta_0$) which satisfy
\begin{equation}
\sum_{i=1}^r \frac{v_i^pv_i^q}{k_i+2} = Z\ \ \ \forall
\ p,\ q\ .
\label{eq:basis}
\end{equation}
Again the non-R symmetries have charges $q$ which differ only by
multiples of $2(k+2)$ or $k+2$ from the $\bar{q}$. These satisfy
the same conditions given in
eqs.(\ref{nur})-(\ref{last}) as were found for the left sector.

In the case of twisted models the Gepner construction is slightly
more complicated. Instead of eq.~(\ref{lr}) we have
\begin{equation}
q_i^J=\bar{q}_i^J-n-2pt_i\ \ \ ,\ \ \
s_i^J=\bar{s}_i^J-n-2m_i\ \ ,
\label{twist}
\end{equation}
where $t_i$ is the twistvector \cite{gepner}. The twist must
commute with supersymmetry (generated by $\beta_0$) and so $t_i$
must satisfy
\begin{equation}
\ \sum^r_{i=1}\frac{t_i}{k_i+2}\ \in\ Z\ .
\label{susy}
\end{equation}

The residual discrete symmetries must commute with the twist and
hence for them the connection between the $\bar{q}$ and $q$
charges does not involve the new term proportional to $t_i$ in
eq.(\ref{twist}). As a result the analysis of the cancellation
conditions is unchanged from the untwisted case and the non-R
symmetries satisfy eqs.(\ref{nur})-(\ref{last}).

In the case that there are twisted sectors the discrete
symmetries are enlarged; the twisted states labeled by $p=0,
\cdots ,(Q-1)$
in eq.(\ref{twist}) transform as $\alpha^p$ under a $Z_Q$ group.
It is straightforward to show that this discrete group satisfies
the same cancellation conditions as the other discrete
symmetries. We start with the orthogonal basis,
eq.(\ref{eq:basis}), of charges including $t_i$ and $\beta_0$.
Taking a linear combination of the anomaly cancellation
conditions of eq.(\ref{nur}) allows us to derive the result
\begin{equation}
\sum_{J=1}^M \Sigma_{i=1}^r t_i(q_i^J+2 p^J t_i)/(k_i+2)=0 \;\
mod\ Z/2\ .
\label{eq:tw1}
\end{equation}
For a twisted model the states have to obey
\begin{equation}
\Sigma_{i=1}^r t_i(q_i+\bar{q}_i+2p t_i)/(k_i+2)=0 \;\ mod\ 2Z\ ,
\label{eq:tw2}
\end{equation}
using the fact that $t_i$ commutes with $\beta_0$ and $\beta_i$
this implies
\begin{equation}
\Sigma_{i=1}^r t_i(q_i+p t_i)/(k_i+2)=0 \;\ mod\ Z\ .
\label{eq:tw3}
\end{equation}
Inserted in eq.(\ref{eq:tw1}) we now obtain
\begin{equation}
\sum_{J=1}^M \Sigma_{i=1}^r \frac{t_i t_i}{k_i+2}p^J=0 \;\
mod\ Z/2\ .
\label{eq:tw4}
\end{equation}
Finally, if we denote by $<k+2>$ the smallest common multiple of
the $(k_i+2)$ terms entering in eq.(\ref{eq:tw4}), we obtain the
required cancellation condition
\begin{equation}
\sum_{J=1}^M\Sigma_{i=1}^r t_i t_i <k_i+2>p^J=0 \;\
mod\ Q^{\prime}/2
\label{eq:tw5}
\end{equation}
where we have written $(k_i+2)<k_i+2>=<k+2>$ and
$Q^{\prime}=<k+2>$ is the discrete group factor left in the
orthogonal basis commuting with supersymmetry. Eq.(\ref{eq:tw5})
is the mixed gravitational discrete symmetry cancellation
condition eq.(\ref{nur}) for the new symmetry associated with
twists. In a similar manner we may derive the remaining
cancellation conditions of eqs.(\ref{eq:u13})-(\ref{last}).

The cancellation conditions for the superconformal models have
immediate implications for the symmetries found in the associated
Calabi-Yau theories. For the $D_{odd}$, $E_6$ or $D_{even}$, $E_7,\; E_8$
modular invariants,  the linear combinations of charges
$\bar{q}_i+q_i$ generate a $Z_{2(k_i+2)}\otimes
\tilde{Z}_{2(k_i+2)}$ or $Z_{(k_i+2)}\otimes \tilde{Z}_{(k_i+2)}$
group respectively. However the non-R symmetries which commute
with $\beta_0$ in the right sector have charges  $\bar{q}+q$
just twice the $\bar{q}$ charges while the combinations $q-
\bar{q}$ vanish. Half the sum of the charges are the Gepner
charges corresponding to the symmetry of the associated Calabi
Yau theory and they satisfy the cancellation conditions
eqs.(\ref{nur}) to (\ref{last}).

\bigskip

In summary we have shown how the anomaly cancellation conditions
of the underlying $U(1)$s of the Gepner construction lead to
conditions on all the non-R discrete symmetries of the model, and
not just those which are subgroups of the $U(1)$s. In many cases
these conditions coincide with the discrete anomaly cancellation
conditions which follow if the discrete symmetry is a subgroup
of an underlying gauge symmetry. However we have not been able
to prove these conditions in all cases (although we have not
constructed any explicit counterexamples). It would certainly be
interesting if the stronger conditions could be proved in general
for this would strongly support the suggestion that all string
symmetries are gauge symmetries.

The cancellation conditions we have found provide constraints on
the structure of any effective low energy theory which can result
from such four dimensional string theories and the related Calabi
Yau theories. These constraints are quite non-trivial and can
restrict the forms of possible symmetries capable of stabilising
the proton\cite{ibanez} and of generating structure in the
fermion mass matrix \cite{scheich}. As such they represent a step
forward in deriving general features following from an underlying
stage of superstring compactification which may allow us to
explore some of the phenomenological implications without relying
on a choice of one out of the many string vacua that are now
known to exist.


\newpage \begin{thebibliography}{99}
\newcommand{\npb}{\mbox{Nucl.Phys.B }}
\newcommand{\pl}{\mbox{Phys.Lett.B }}


\bibitem{ibanez} L.E. Ibanez, G.G. Ross, \pl 260 (1991) 291
\bibitem{priv} Anomaly cancellation conditions in string theories
have also been studied for a different class of discrete
symmetries by L.E. Ibanez and D.L\"ust, \npb 382 (1992) 305
\bibitem{ross+} L.E. Ibanez, G.G. Ross, \npb 368 (1992) 3
\bibitem{krauss} T. Banks, Santa Cruz preprint SCIPP 89/17
(1989);                 L. Krauss, F. Wilczek, Phys.Rev.Lett. 62
(1989) 1221;                  T. Banks, \npb 323 (1989) 90;
                 L. Krauss, Gen.Rel.Grav. 22 (1990) 50;
                 M. Alford, J. March-Russell, F. Wilczek,
                 Nucl.Phys.B 337 (1990) 695
\bibitem{ade} A. Capelli, C. Itzykson, J.G. Zuber,
Commun.Math.Phys. 113               (1987) 1
\bibitem{kopp} V.A. Fateev, A.B. Zamolodchikov, Sov.Phys. JETP
62(2),                August 1985
\bibitem{andy}C.A. L\"utken, G.G. Ross, \pl 214 (1988) 357
\bibitem{cordes}           S.F. Cordes, Y. Kikuchi, Mod. Phys.
Lett. A 4               (1989) 1365
\bibitem{gepner} D. Gepner, \npb 296 (1988)
757; \pl 199 (1987) 380;                  Princeton preprint,
December 1987
\bibitem{vafa} B.R. Greene, C. Vafa, N.P. Warner, \npb 324 (1989)
371
\bibitem{wir} J. Fuchs, A. Klemm, C. Scheich, M.G. Schmidt, Ann.
of Phys.               204 (1990) 1
\bibitem{lt} L.E. Ibanez, private communication
\bibitem{ibanezneu} L.E. Ibanez, CERN Preprint CERN-TH-6662-92
\bibitem{scheich}G.G. Ross and C. Scheich, in preparation

\end{thebibliography}
\end{document}